# Aboriginal Oral Traditions of Australian Impact Craters

Duane W. Hamacher[1] and John Goldsmith[2]

[1]Nura Gili Centre for Indigenous Programs, University of New South Wales, Sydney, NSW, 2052, Australia
Email: d.hamacher@unsw.edu.au

[2]Curtin Institute for Radio Astronomy, Curtin University, GPO Box U1987, Perth, WA, 6845, Australia
Email: john.goldsmith@icrar.org

**Abstract**

We explore Aboriginal oral traditions that relate to Australian meteorite craters. Using the literature, first-hand ethnographic records, and fieldtrip data, we identify oral traditions and artworks associated with four impact sites: Gosses Bluff, Henbury, Liverpool, and Wolfe Creek. Oral traditions describe impact origins for Gosse's Bluff, Henbury, and Wolfe Creek craters and non-impact origins of Liverpool crater, with Henbury and Wolfe Creek stories having both impact and non-impact origins. Three impact sites that are believed to have formed during human habitation of Australia - Dalgaranga, Veevers, and Boxhole - do not have associated oral traditions that are reported in the literature.

**Notice to Aboriginal and Torres Strait Islander Readers:** This paper may contain the names and images of people that have died.

1. Introduction

Prior to European colonization by the British in 1788, Aboriginal Australians were a predominantly hunter/gatherer people with several hundred distinct languages and dialects spanning all corners of the continent. Genetic and archaeological evidence shows that Aboriginal people migrated to Australia from South East Asia over 40,000 years ago (Bowler et al., 2003), with upper limits exceeding 60,000 years (Rasmussen et al., 2011). Aboriginal cultures did not develop written languages, relying instead on strong oral traditions where knowledge was passed through successive generations in the form of story, song, dance, art, and material culture. Aboriginal knowledge is typically transmitted through the "Dreaming", which is the embodiment of the oral traditions, laws, customs, and culture of the community (Rose, 1992). In Aboriginal cultures, the Dreaming is sometimes thought of as a time in the distant past where spiritual ancestors and beings formed the land and sky. It can also be thought of as a current parallel reality and is generally considered non-linear in time.

Aboriginal knowledge systems include explanations about the natural world, which have practical applications and are used for predictive purposes. This research area is known as "ethnoscience". Ethnoscience relates to environmental and atmospheric science (Clarke, 2009), astronomy (Clarke, 1997; Norris and Hamacher, 2009), ecology (Vigilante, 2004), botany (Clarke, 2007), zoology (Isaacs, 1996), and geography (Walsh, 1990). Aboriginal knowledge regarding meteoritics is evident in Aboriginal traditions of impact events not associated with known impact sites (Hamacher and Norris, 2009), meteorites (Bevan and Bindon, 1996), meteors (Hamacher and Norris, 2010), and comets (Hamacher and Norris, 2011). The hypothesis that Aboriginal people witnessed impact events and recorded them in their oral traditions is a form of "ethnogeology" - explanations regarding the formation and nature of geological features (Murray, 1997). The ethnogeology of Wolfe Creek crater has been explored (e.g. Goldsmith, 2000; Reeves-Sanday, 2007) as well as impact events described in Aboriginal oral traditions currently unknown to Western science (Hamacher and Norris, 2009).

Of the 26 confirmed craters in Australia (Bevan and McNamara, 2009), the six smallest craters, Boxhole, Dalgaranga, Henbury, Liverpool, Veevers, and Wolfe Creek, each with diameters < 2 km, are also the youngest (see Figure 10 for locations of all places described in this paper). Each of these craters is identifiable as a prominent, unusual feature in the landscape. Four of them are believed to have formed within the Holocene era, only 25% of the minimum time humans have inhabited Australia: Henbury (≤ 4,700 years), Boxhole (5400±1500 years), and possibly Veevers (< 4,000 years) and Dalgaranga (< 3,000 years) (Haines, 2005). Because of this, we search for records of these six impact sites in Aboriginal oral traditions. Many of the larger craters are either buried or eroded to such a point that they are not easily distinguishable from the surrounding landscape and are geologically very old. Therefore, we do not expect to find stories associated with these structures. Some of the larger, older craters are more prominent on the landscape, such as Gosses Bluff (Northern Territory) or Goat Paddock (Western Australia), or possess unusual features, such as Spider crater (Western Australia). Other craters may form bodies of water, such as Lake Acraman (South Australia) or Shoemaker crater (formerly Teague, Western Australia). We might expect to find oral traditions associated with such structures but do not expect them to relate to a cosmic impact. We also suspect that oral traditions associated with some craters exist, but simply have not been collected or published.

In this paper, we explore Aboriginal traditions relating to Australian impact craters and seek to find out if these traditions describe craters as originating from a cosmic impact. Data used in this paper were collected from ethnographic fieldwork, published ethnographies, historical and ethno-historical documents, linguistic material, and various Aboriginal artworks. This paper represents a treatise of Aboriginal traditions regarding confirmed Australian impact craters and provides new information that has not been reported in the literature.

In the following sections, we describe traditional knowledge regarding Gosses Bluff, Henbury, Liverpool, and Wolfe Creek craters but find none associated with Boxhole, Dalgaranga, or Veevers craters. We are faced with several possibilities to explain the presence or absence of these stories:

1. The story is based on a witnessed event and was recorded in oral traditions;
2. The formation of the crater was not witnessed, but was instead deduced and incorporated in oral traditions;
3. The formation of the crater was not witnessed, and stories explaining it as an impact site are coincidental;
4. The origin or nature of the crater is not part of a structured oral tradition (Dreaming), but is generically attributed to supernatural elements or grouped in with general landscape features;
5. Impact stories were influenced by Western science;
6. Related stories may have once existed but have been lost for whatever reason;
7. No stories of the crater ever existed.

It is difficult to know which possibility is true in each instance, but we explore each of these with reference to the craters described above in the following sections.

## 2. Craters With Known Oral Traditions

### 2.1 Gosses Bluff Crater (*Tnorala*)

Gosses Bluff is a highly eroded complex structure measuring 22 km in diameter with the central uplift forming a ring-shaped mountain range measuring 5 km wide by 250 m high (Milton et al., 1996). The crater, which is 160 km west of Alice Springs, formed from an impact 142.5±0.8 million years ago (Milton and Sutter, 1987) and was identified by Edmund

Gosse in 1873 (Crook and Cook, 1966). The structure was mapped by the Bureau of Mineral Resources in 1956 and was considered to have formed by either a volcanic or an impact event (e.g. Crook and Cook, 1966; Cook, 1968). It was not until the early 1970s that it was generally accepted as an impact structure (Milton et al., 1972).

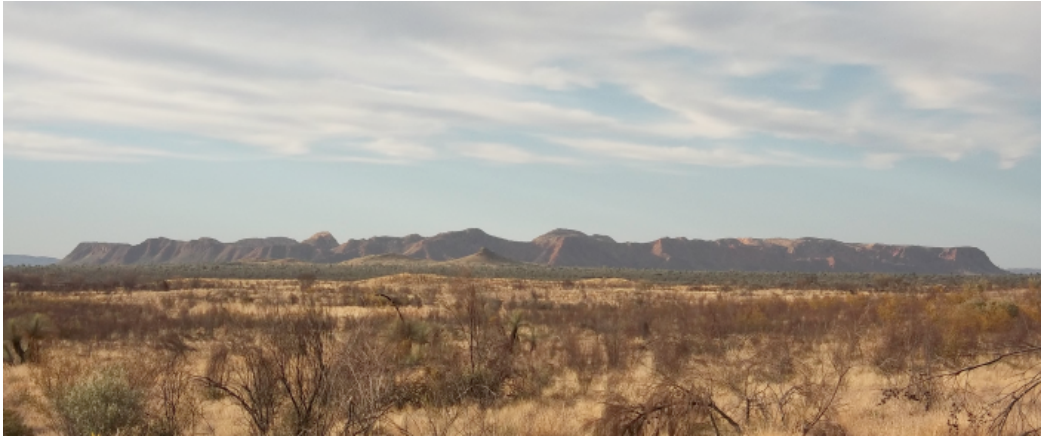

*Figure 1. Gosses Bluff (Tnorala) as seen from the south. Photo by D.W. Hamacher.*

Gosses Bluff, called *Tnorala* in the Western Arrernte language (Figure 1), is registered as a sacred site. Herman Malbunka was the custodian of the Tnorala story. Upon his death, his wife, Mavis, became the caretaker (*kurturngula*) of the story. The story of Tnorala is now well known due to documentaries and film dramatisations, mainly for the fact that the Aboriginal story mirrors the scientific explanation of the crater. According to Mavis Malbunka (Thornton, 2007), in the Dreaming a group of eight women took the form of stars and danced in a corroboree (ceremony) in the Milky Way. One of the women was carrying a baby and placed him in a *coolamon* (or *turna* - a type of wooden basket) and set him on the Milky Way. As the woman went back to continue dancing with the others, the turna slipped off the Milky Way. The baby struck the earth and was covered by the turna, the force of which drove the rocks upward, forming the mountain range we see today. The baby's mother, the evening star, and father, the morning star, continue to search for their baby to this day (see also Parks and Wildlife Commission of the Northern Territory, 1997:1; Cauchi, 2003).

Malbunka continues:

> *"We tell the children don't look at the evening star or the morning star, they will make you sick because these two stars are still looking for their little baby that they lost during the dance up there in the sky, the way our women are still dancing."*

Malbunka never specified Venus as the morning or evening star, although it is assumed. She notes that the morning and evening star are not always visible ("*They don't show themselves all the time. No! Only every now and then*"), consistent with a reference to Venus. But instead of Venus, Malbunka identified the "morning" and "evening" stars as a mysterious phenomenon known as a "min-min lights." These "ghost lights" are prominent in the Aboriginal folklore of eastern Australia (Pettigrew, 2003). She recounted an anecdote about two health workers in Hermmansburg (Ntaria) that were pursued by a min-min light, which she claimed was the baby's mother searching for her lost child. It is possible that the story of the min-min light was imported from communities further east and incorporated into the current oral traditions.

Malbunka states that the celestial turna *"…is still there. It shows up every winter."* Malbunka did not identify the celestial turna in the media that feature the Tnorala story. During the Meteoritical Society 2012 fieldtrip to the Central Desert, the tour-guide pointed out the turna in the sky as the arch of stars in the constellation Corona Australis (the Southern Crown). The arch of stars forms the same shape as a turna seen from the front (Figure 2) and is visible south of the galactic bulge high in the winter night sky after sunset. The guide claimed to learn the identity of the celestial turna from the local community, but it is not reported in the published literature or discussed in ethnographic field notes.

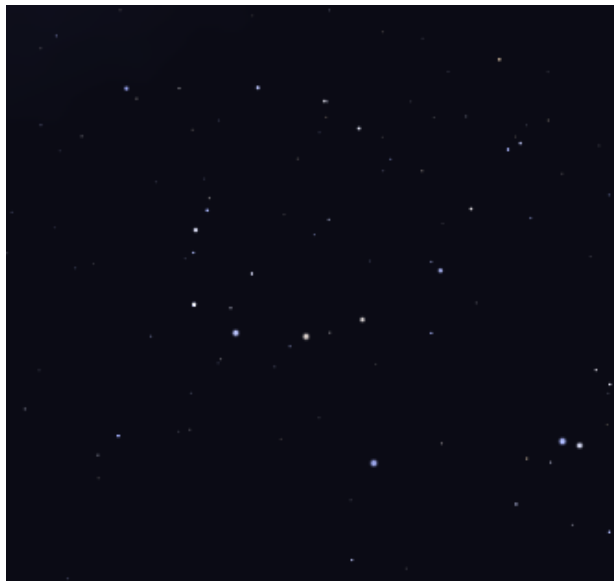

*Figure 2. The constellation Corona Australis. The curve of stars is seen as the turna falling from the Milky Way. Image from Stellarium software package (www.stellarium.org).*

The story is visualized in an Aboriginal painting by Lynette Spencer, which is featured on the bottom of tourist signs at the site (Figure 3). The meanings of the motifs in the painting are not provided, although the anthropomorphic figures featured at the top of the sign are based on Arnhem Land figures. These are not part of the painting by Spencer.

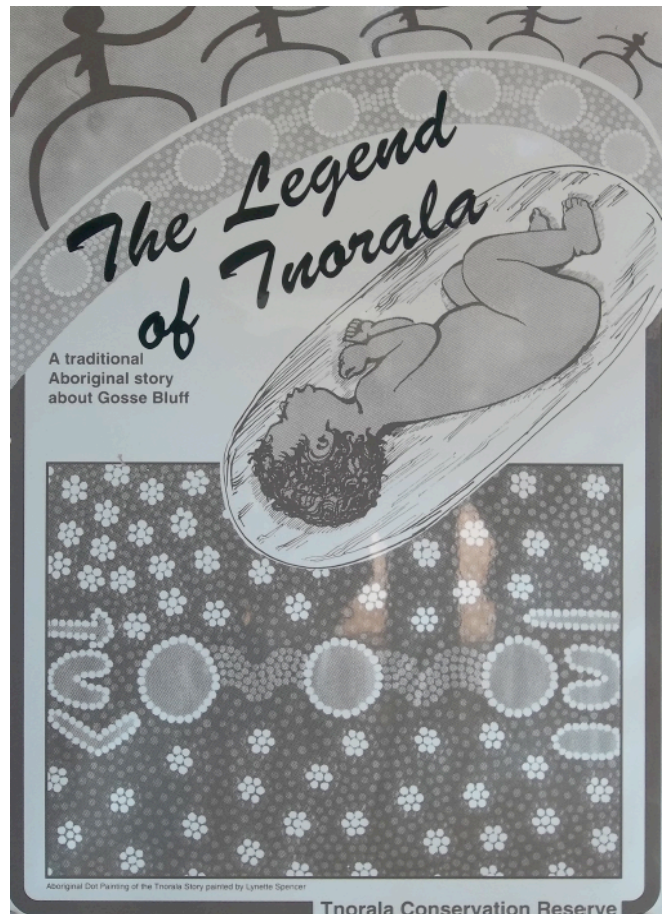

*Figure 3. Information sign at the Tnorala Conservation Reserve. Painting by Lynette Spencer. Photo of painting by D.W. Hamacher.*

According to Arrernte traditions (as noted on signs at the site):

> "*Tnorala was a special place where many bush foods were created by rubbing special rocks made by the falling star, the Milky Way baby. On the baby's body were many sacred stories, about other matters of importance, which were held in these rocks. Men were responsible for some rocks and women for others. Each rock represented a particular bush food. Men or women would rub the rock and sing, so that there would be a good supply of the food that rock represented*".

The signs claim that the sacred 'rubbing rock' has gone missing. Because of this, a food called *puraltja* no longer grows in large quantities, except after heavy rain and Malbunka asks visitors not to take rocks from Tnorala. There is no information regarding the identity of these rocks, but they could not be remnant fragments of the impactor.

Local traditions describe the center of Tnorala as the site of a massacre. The story of the massacre does not appear in the anthropological literature, but is provided on signs posted in the reserve based on local traditional knowledge. The story tells that long before Europeans came to Australia, a community lived in Tnorala. One day one of the men went hunting for kangaroo. When he returned, he found the all of his people had been murdered, including women and children. He knew this was the act of the Kurdaitcha men – fierce warriors who lived in the desert to the south. The man informed the rest of his family, who lived in the ranges nearby. They formed a party and hunted down and killed the Kurdaitcha men. Because of the massacre, Tnorala is known as a "sorry place" and the center of the crater is considered sacred ground. It is not clear if this is story is based on an actual event or is mythological in nature.

Nothing is reported in the anthropological literature about the Dreaming story of Tnorala. It only appears in media reports and local knowledge. Four possibilities exist regarding the origin and nature of this Dreaming story:

1. The story pre-dates colonisation but the impact story is coincidental to Tnorala being an impact crater,
2. The story pre-dates colonisation but the impact origin was deduced by Aboriginal people based on the structure's morphology (it makes the shape of a crater similar to dropping a heavy stone in sand),
3. The story is post-colonisation and was influenced by scientists during the last several decades of intense research at the crater, OR
4. The story pre-dates colonisation but changed to incorporate new knowledge gained from Western science.

In the fourth case, the term "pre-colonisation" could be replaced with "before European interest in the site", even if the story developed after colonisation. Malbunka does not say how old the story is and it is not known how many generations through which this story was passed. The nature of Aboriginal traditions is such that stories do not necessarily have an "origin" in linear time. New events may be incorporated into existing oral traditions or may form the basis of a new oral tradition (see Hamacher and Norris, 2009). One should be warned that asking an Aboriginal person about the "date of origin" of a story may be inappropriate, as the Dreaming in many Aboriginal cultures still exists and is not restricted to a time in the distant the past.

It is probable that scientists researching the site had some influence on local traditions, but the degree of this influence is unknown and we cannot assume that impact stories associated with Gosses Bluff *must* have originated from Western science. A number of impact stories can be found across Australia with no associated impact structures or meteorite falls known to Western science (Hamacher and Norris, 2009) - some of which are from Arrernte country. A survey of meteoritic traditions might help us understand the story of Tnorala. In some Arrernte traditions, meteors signify that the spirit of a person who died far from their home was returning to their home country (Basedow, 1925:296). They could also represent the presence of an evil magic called *Arungquilta* (Spencer and Gillen, 1899: 566). Other Arrernte traditions identify meteors as the fiery eyes of celestial serpents that drop into deep waterholes (Strehlow, 1907: 30). A story recorded by Róheim (1945: 183) highlighting this view comes from Palm Valley in the nearby MacDonnell Ranges. In the story, a star fell into a water hole where the serpent *Kulaia* lived, making a noise like thunder. Western Arrernte traditions claim that first human couple originated from a pair of stones that were thrown from the sky by the spirit Arbmaburinga (Róheim, 1971: 370).

Unfortunately, none of these stories shed light on the Tnorala story, as it did not involve serpents, evil magic, or death. It only involved a lost baby and the parents searching for their child. The Tnorala story does not explicitly say the baby died, only that it was covered by the turna. Future work with Arrernte custodians may help clarify these points.

## 2.2 Henbury Crater Field (*Tatyeye Kepmwere*)

The Henbury crater field (Figure 4) is approximately 130 km south of Alice Springs and comprises 13-14 simple impact craters spread out over a square kilometer (Alderman, 1932; Haines, 2005). The craters, ranging in size from 180 m to < 10 m in diameter, were formed from the impact of a nickel-iron meteoroid that fragmented in the atmosphere ≤ 4,700 years ago (Kohman and Goel, 1963). Given the relatively young age of the impact, one might expect the event is recorded in Aboriginal oral traditions.

The craters lie on Henbury station, which is 5,273 km$^2$ in total size, was operated on a pastoral lease from 1877, founded by Walter and Edmund Parke (Buhl and McColl, 2012:13). Walter identified the craters in 1899 and wrote to the anthropologist Francis Gillen, describing it as *"one of the most curious spots I have ever seen in the country."* Parke was unsure of structures' origin and stated, "*To look at it I cannot but think it has been done by human agency, but when or why, goodness knows!*" (Buhl and McColl, 2012; Parks and Wildlife Commission of the Northern Territory, 2002: 2).

In January 1931, 33 years after Walter Parke found the craters on Henbury station, a prospector named James Maxwell Mitchell sent a meteorite fragment to the University of Adelaide. Through a series of events, it came to the attention of a young geologist named Arthur Alderman, who conducted the first scientific investigation of the site a few months later in May 1931 (Alderman, 1932). In an addendum in Alderman's paper, L.J. Spencer included communication with Mitchell who had visited the site some years earlier and was the first person on record to identify it as a meteorite impact site (see Buhl and McColl (2012) for a full treatise of the history of Henbury research).

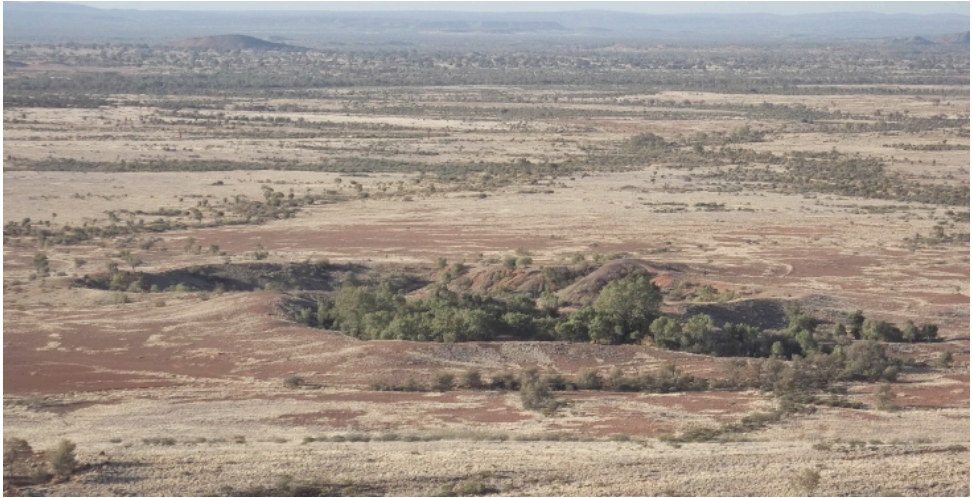

*Figure 4. The larger cluster of the Henbury craters as seen from the Bacon Range looking to the northeast. Photo by D.W. Hamacher.*

According to James S. M. Mitchell (1934), he visited a blacksmith in Todmorden in 1916 when he noticed a peculiar piece of metallic iron that he believed contained nickel. When he learned that the object had been taken from the "blowholes" at the Henbury cattle station, he visited the station and found a number of large metallic stones around the craters and concluded that they had "*dropped from a molten mass falling at great speed*" and that "*huge masses of metal probably lay buried in the bottom of the craters.*" Unfortunately, Mitchell passed away the following year in 1935. His obituary (Anon, 1935) claims that while he was on a prospecting expedition in 1921, his attention was further drawn to the craters after his Aboriginal guide refused to go near them. According to the obituary, local Aboriginal traditions described the place as where a "fire debil-debil came out of the sky and killed everything in the vicinity."

Alderman's study confirmed the craters were of meteoritic origin. The discovery of a well-preserved crater field surrounded by meteorite fragments was of worldwide interest and in July 1931, newspapers across the globe reported the find (e.g. Anon 1931a,b). During this period, reports clearly stated that Aboriginal people in the district "have no legends or stories regarding the place, nor do they appear particularly interested in it" (Anon, 1931a). Sometime between July and November 1931, Mitchell took an elder Aboriginal man to the site, but the man would not venture "within half a mile" of any crater or "camp within two miles of them", describing them as "*chindu chinna waru chingi yabu*", literally meaning "*sun walk fire devil rock*". The Aboriginal man expanded upon the meaning of the name and told Mitchell that a fire-devil (*chinka waroo*) lived in the rock-hole (*yabo*). He claimed his paternal grandfather had seen the fire-devil and that it came from the sun. He also said that Aboriginal people were not to drink water that collected in the bottoms of the larger craters, as they feared the fire-devil would "*fill them with a piece of iron*." The Aboriginal man warned Mitchell not to go near them and told him that the only reason the fire-devil did not attack him during daylight was because he was "grey-headed". This affirms Mitchell's earlier encounter in 1921 with a local Aboriginal guide and his refusal to approach the Henbury craters. Anon (1945) claims that while the craters were being studied by Alderman in 1931, the custodian of the Aboriginal traditions of the Henbury area was nowhere to be found, having decamped elsewhere. This would not be an unexpected reaction if the custodian felt the area was taboo. The article stated that this view of the craters was held across the Petermann Ranges and surrounding areas (Mitchell, 1934). News reports with headlines reading "SUN WALK FIRE DEVIL ROCK" (Anon, 1931b) appeared, promoting the new find and suggesting the craters' formation was fairly recent in geologic history.

Prof. J.B. Cleland was going to investigate stories about the craters (Anon 1932b), but nothing is found in the literature that report on his findings. In March 1932, a local resident of Kadina undertook an independent investigation of the Henbury craters (Anon, 1932a). The resident (name not provided) claimed that he and his friend contacted the Aboriginal "doctors" or "wise men" (elders) from the "Western tribes" to learn more about their perspective of the crater field. According to the resident's Aboriginal contact, all young [Aboriginal] men and the women were forbidden from approaching the craters. The Aboriginal contact said of the place "*Schindo waroo chinka yabbo shinna kadicha cooka,*" which he translated to "a fiery devil ran down from the sun and made his home in the earth. He will burn and eat any bad blackfellows" (Anon 1932b). [Note 1]

This account is interesting for two reasons. First, it clearly suggests a living memory of the Henbury impact. Second, the destructive event was seen as divine punishment. Such disasters are often attributed to people breaking laws and taboos. In a similar vein, the Indigenous Moki (Hopi) people of Arizona in the United States recounted an oral tradition about a "blazing star which fell years ago, when the oldest of the ancient cliff dwellings was new," a place called Meteor Mountain (Anon, 1912), known today as Meteor (Barringer) Crater, which formed 50,000 years ago – long before humans are believed to have settled the New World (e.g. Fagundes et al. 2008). According to Hopi traditions, they had "offended a Great Spirit, and this blazing star had come as a warning, lighting up the earth for hundreds of miles around, and spreading terror throughout the repentant tribes," (Anon, 1912). It is uncertain if Western scientists influenced the story or if the meteorite impact was conflated with more recent volcanic eruptions nearby about a thousand years BP (Malotki 1987). In either case, the Hopi traditions describe the impact as divine punishment for unrepentant, or "bad", people. Similar accounts from Australia relating meteors and punishment are evident throughout Australia (Hamacher and Norris 2009, 2010).

According to Spencer and Gillen (1904: 28), Aboriginal people that camped near the Finke (Larapinta) River at Henbury Station were Arrernte speakers but called their camp at Henbury *Waingakama.* The Finke River lies 7 km north of the crater field. Further study shows that the Henbury craters lie within country crossed by multiple Aboriginal language/dialect groups, including Arrernte, Luritja, Pitjantjatjarra, and Yankunytjatjara. An investigation of the Aboriginal words describing the Henbury craters recorded by Mitchell and the Kadina resident reveals that they are from the Luritja language (Hansen and Hansen, 1992). Luritja is a dialect of the Western Desert language that shares close similarities with Pintupi, Pitjantjatjarra, and Yankunytjatjara (Goddard, 1992). The identity of the words cited by Mitchell and the Kadina resident in the Luritja language are as follows using Hansen and Hansen (1992): "*chindu*" or "*schindo*" (*tjintu*) refers to the sun (p. 145); "*chinna*" or "*shinna*" (*tjina*) refers to feet or footprints (p. 145) but can also indicate an foot action like walking or running; "*waroo*" (*waru*) refers to fire or heat (p. 171); "*chinka*" (*tjinka*) is a word used in various Western Desert languages meaning "dead" or "devil" (David Nash, per. comm.), "*yabu*" or "*yabbo*" (*yapu*) refers to a rock or hill (p. 188), "*cooka*" (*kuka*) refers to meat or eating meat (p. 31). The word Kadicha (*Kurdaitcha*) is common among Central and Western desert groups, including the nearby Arrernte and Warlpiri. The term has multiple but similar meanings, generally referring to a spirit that punishes evildoers (e.g. Spencer and Gillen 1899: 476).

Interest in the Henbury craters and associated meteorites led to a demand for fragments of the Henbury impact. By 1945, Aboriginal people in the region had taken note of this demand and began selling "pieces of the star that fell from the sky" (Vox, 1945). It was not uncommon for Aboriginal people to recognise and take advantage of the demand for meteorites and tektites. Aboriginal people in the Western Desert often collected tektites to sell to white prospectors (which the Aboriginal people colloquially called "meteorites") until the demand waned and the specimens were lost or simply discarded (Hamacher & O'Neill, 2013). There is little doubt that public and scientific interest in the craters and meteorites had an influence on the local Aboriginal people, but it is unknown what influence this had on their traditions.

Unlike Mitchell or the Kadina resident, Alderman was unable to find any stories about the site from the local Aboriginal community. He concluded that the Aboriginal people seemed to have "no interest" in the Henbury craters or any ideas as to their origin. Despite this information about Aboriginal traditions of the crater field being published, news accounts for many years continued to make the claim that no Aboriginal traditions existed regarding the craters (e.g. Anon, 1934). There are a number of possible explanations for Alderman's inability to obtain any oral traditions of the site. Aboriginal people did not often share their information or stories with white people unless a trust and rapport had been developed first. It was not uncommon for a stranger to ask an Aboriginal person about a story or place and be given a dismissive response or one of feigned interest, while a more trusted white person would be given the full account.

The Aboriginal Areas Protection Authority advises that a sacred site is recorded from the center of the crater field, which is called *Tatyeye Kepmwere* (*Tatjakapara*) in the Arrernte language (Parks and Wildlife Commission of the Northern Territory, 2002: 15). The commission reports that stories of the site are known but will be "*used for interpretation purposes after agreement by the Aboriginal custodians of the site.*"

The only story recorded in the literature is from Mountford (1976: 259-260) [Note 2], which relates to the largest crater only. In the story, Mulumura (a lizard-woman) was camping within the largest crater. The woman picked up handfuls of soil and tossed them away, creating the structure's bowl shape. The discarded soil formed the ejecta rays that were once visible at the site. The rays at Henbury are unique for terrestrial impacts and closely resemble ejecta rays found on lunar craters (Fortowski et al., 1988; Milton and Michel, 1977). Unfortunately, traffic and prospecting at the site has almost completely destroyed the rays (Buhl and McColl, 2012). The story also mentions piles of meteoritic iron that were once in the crater but were tossed out along with the soil.

Mountford (1976: 260) provides a map detailing the various elements of the story, which is provided here (Figure 5). In the figure, **A** was the camp of Mulumura, **P** was where she slept; **C** was her windbreak; **B** represents the soil she picked up and discarded from the crater forming the ejecta rays that were once visible at the site. It also includes piles of meteoritic iron. **R** represents her wind-break, and the acacia tree; **D** represents the water hole on the Finke River; **F** is the Henbury cattle station; **G** is a low hill covered with acacia trees; **E** is the Finke River; and **N** is a creek emptying into the river. To the south of the craters is a ridge, called Bacon Ridge, which is denoted by **H**. Sacred ceremonial objects in Arrernte culture, called *Tjurungas*, were kept in a small cave **J** near the top of this ridge; **K** is one of the stone Tjurungas; **L L L** represent the long wooden Tjurungas that were placed in the cave at the conclusion of ceremonies; and **M** indicates marks of blood that were poured on the walls when the objects were stored in the cave. Mountford was unable to obtain stories about the other craters, but claimed that "*there is little doubt that such myths exist*" (p. 260). Mountford (1976: 259) cites this story as evidence that the Aboriginal people have no living memory of the meteorite fall. It should be noted that Mulumura is prominent in Pitjantjatjarra traditions related to *Kata Juta*, a large rock formation also known as The Olgas located 250 km west-southwest of Henbury (Mountford, 1976: 488).

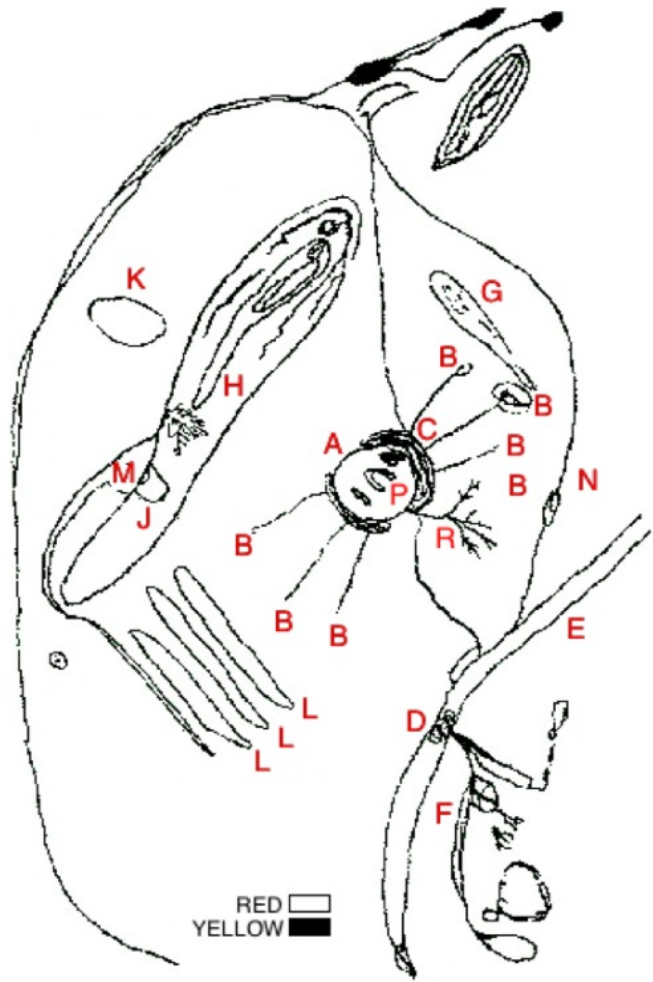

*Figure 5: A map of the Mulumura story from Henbury (after Mountford 1976: 260).*

Mountford's record of the Mulumura story probably relates to Aboriginal oral traditions of the Finke River within Henbury station about a great totemic lizard god. Worms (1952: 554) translates passages from Strehlow (1907), a German Lutheran missionary who documented the Arrernte and Luritja languages: "In Henbury at the Finke River, the *tjunba* (*Varanus giganteus* grey - the largest species of monitor lizards in Australia) men lived ages ago. In a plain near Henbury, where many eternal men and women were living, the inhabitants of this camping place had already been instructed by *tjunba* (i.e. the great Lizard and totem-god). Southwest of Hermannsburg, there is the great totem place, called *Manjiri*. Once many lizard-men (*atna tjunba*) were living here."

The local Luritja people recognized the Henbury craters as a place where an unusual event involving fire and destruction occurred. This recognition, combined with a fear that the fire-devil, which came down from the sun and made his home in the craters and would fill them with iron if people ventured too close, indicates either a living memory of the event or an acknowledgement of how the craters were formed. But one must use caution when interpreting this translation in the absence of context. It is also possible that the Henbury craters were generically attributed to spirits or spirit activities ("devils") to the Luritja, while other language groups incorporated the site as part of the Mulumura Dreaming story.

Bevan and Bindon (1996: 99) suggest that Western knowledge of the site influenced the Aboriginal account noted by Mitchell. This is certainly possible, as the meteoritic nature of Henbury was possibly known to locals (or at least the idea was discussed) prior to Mitchell's confirmation of Henbury as an impact site. There are other examples of Western knowledge influencing crater traditions, as we will see with Wolfe Creek crater in Section 2.4.

It is also possible that the Henbury impact was witnessed, incorporated into local oral traditions, and dispersed across the country (Bevan and Bindon, 1996). An example comes from Paakanji country near Wilcannia in northwestern New South Wales, 1400 km south east of Henbury (Jones, 1989). According to the story a large fiery star rumbled and smoked as it fell from the sky, crashing to the earth, burning some while killing others. Jones gave the location of the event as a place called *Purli Ngaangkalitji* (meaning "Fallen Star") in the Darling riverbed near Wilcannia. Jones provided a map of Purli Ngaangkalitji, but when Australian National University astronomer and comet-hunter Robert McNaught surveyed the area, he found no evidence of an impact or of meteoritic material (Steel and Snow, 1992: 572). Jones claimed she first heard the story in 1927 and that it was "very old". Other stories from across Australia describe fiery objects falling from the sky, lighting up the surrounding landscape and striking the earth with a deafening roar, sending debris into the air and causing fire, death, and destruction (see Hamacher and Norris, 2009 for a treatise on the subject).

The question is whether or not an oral tradition can survive for thousands of years. As a test of short-term memory regarding a meteoritic event, we examine the indigenous Evenki (Tungus) people of central Siberia, who witnessed a meteoritic airburst in 1908 that destroyed 2,150 square kilometers of forest near Tunguska, Russia. Some 15 years passed before researchers made it to the site and began collecting stories and accounts of the airburst from local villagers and indigenous groups. According to researchers that visited the region (Krinov, 1966: 125-143), memory of the event was still alive in the minds of the local people. Using accounts reported in newspapers shortly after the event, it seems details of the airburst in Evenki folklore changed to the point that numerous discrepancies and contradictions were evident after only 15 years. Because of this, the researchers faced an "insurmountable difficulty in establishing the truth" of the event (*ibid*: 135).

This is not entirely unexpected, as events that took place many years in the past are susceptible to changed and altered memories in the minds of the witnesses. These altered memories could be due to any number of memory errors known to psychologists. An example is *transience* – the loss of memory as time passes (Schacter, 2011: 243). This affects the quality of a memory, the details of which tend to deteriorate from specific to general. Other forms of memory error include *confabulation* – the recollection of inaccurate or false memories (Loftus, 1997); *unconscious transference* – misattribution of the source of a memory (Deffenbacher et al., 2006), *imagination inflation* – details of a memory that are exaggerated in the mind (Mazzoni and Memon, 2003); or *schematic errors* – where a schema (organized pattern of thought) is used to assist in constructing elements of an event that cannot be recalled (Kleider et al., 2008). It is evident that specific and accurate details of an event in memory will deteriorate over time. Therefore, we might not expect to find "accurate" details of a meteorite impact in oral tradition after thousands of years, if at all. General details of an event could remain in memory, although the length of this time is a matter of ongoing debate. Examples of "deep time" oral tradition, which could last for thousands of years (e.g. Henige 2009), have been proposed by a number of authors (e.g. Echo-Hawk 2000; Hamacher & Norris 2009; Piccardi and Masse 2007), but the topic remains one of ongoing debate.

## 2.3 Liverpool Crater (Yingundji)

Liverpool crater in Western Arnhem Land, Northern Territory is a 1.6 km wide simple impact structure that formed during the Neoproterozoic (Haines, 2005). Confirmed as an impact structure in 1970 (Guppy et al., 1971; Haines, 2005), Liverpool crater lies ~2 km north of the Liverpool River and is only accessible via helicopter. The Aboriginal name of the crater is *Yingundji* in the Kunwinjku language (Shoemaker and Shoemaker, 1997).

Upon visiting the site in 1996 to shoot the York Films documentary "Three Minutes to Impact", Eugene and Carolyn Shoemaker were accompanied by a family of Aboriginal men, including the famous artists, brothers Johnny Maurirundjul and Jimmy Njimimjuna of the Kurulk clan (Shoemaker and MacDonald, 2005: 479). While Eugene mapped the structure, Maurirundjul exchanged stories with Carolyn regarding the origin of the structure. He explained that in their traditions the crater was the nest of a giant catfish. Later, on visiting some of the rock shelters in the crater, the Shoemakers found several Aboriginal paintings, one of which showed the giant catfish.

Fish comprise a majority of rock art motifs in Western Arnhem Land and large catfish inhabit the Liverpool river system (Taçon, 1988). Numerous motifs of giant catfish appear in Arnhem Land rock art and the catfish is important in some ceremonies (Elkin, 1954; Taçon, 1988; Paul Taçon personal communication). We were unable to identify any rock art from the Liverpool crater in the literature, but we have plans for ethnographic and archaeological research the site.

### 2.4 Wolfe Creek Crater (Kandimalal)

Wolfe Creek crater (Figure 6) is the largest impact structure in Australia from which meteorite fragments have been found (Haines, 2005: 48). The simple crater is located ~130 km south of Halls Creek on the edge of the Great Sandy Desert in the East Kimberley, Western Australia (Guppy and Matheson, 1950; Bevan and de Laeter, 2002). It is approximately 860 meters in diameter with an estimated age of ~300,000 years (O'Neill and Heine, 2005). Frank Reeves, N.B. Suave, and Dudley Hart were the first Westerners to identify the structure (Reeves and Chalmers, 1949). The crater was sighted in 1947 during an aerial survey of the Canning Basin (Bevan and McNamara, 2009). A.J. Jones, a constable from Halls Creek, claimed that an Aboriginal tracker had shown him the crater in 1935, but the claim remains unsubstantiated (Gard and Gard, 1995).

The crater lies within the traditional lands of the Jaru, who call the structure *Kandimalal* (Tindale, 2005:376). Nearby language groups include the Walmatjarri, Kukatja and the Ngarti (Horton, 1996). In the late 1990's, Peggy Reeves-Sandy, a University of Pennsylvania anthropologist and the daughter of Frank Reeves ("co-discoverer" of Wolfe Creek Crater), visited the crater and conducted ethnographic fieldwork on Aboriginal knowledge relating to the crater and representations of the crater in art (Reeves-Sandy, 2007). The results of her work were presented in an exhibition entitled "Tracks of the Rainbow Serpent" and the book "Aboriginal Paintings of the Wolfe Creek Crater, Track of the Rainbow Serpent" (*ibid*). Other researchers have collected cultural knowledge of Kandimalal since 1947, including Norman Tindale in 1953 (Tindale, 2005: 376) and the co-author (Goldsmith) between 1998 and 2011 (Goldsmith, 2000). An extensive investigation into Aboriginal astronomical knowledge and belief relating to Kandimalal is the subject of Goldsmith (2013).

Early accounts state that Kandimalal has "*no particular meaning in their language, and no legend exists to hint of its origin*" (Cassidy, 1954), a claim reminiscent of the Henbury craters. However, Aboriginal communities continue to maintain various stories associated with Kandimalal (sometimes spelt Gandimalal). One of the stories tells of a pair of subterranean Rainbow Serpents that created the nearby Wolfe and Sturt creeks. One serpent emerged from the ground, creating the circular structure of Kandimalal (Bevan and McNamara, 2007; Goldsmith, 2000; Reeves-Sanday, 2007). Similar Dreaming stories are found in the Western Desert, which typically involve a pair of ancestral snakes burrowing under the Earth, forming rivers and emerging from the ground to form rockholes and claypans (Graham, 2003: 32). Crater features are incorporated in the oral traditions. For example, in some Jaru traditions, the head of the serpent formed a depression on the southeast rim of the crater as it emerged from the ground (Reeves-Sanday, 2007: 99). Other Jaru artists claim that the crater was formed by an "Old Fellow" digging for yams. They say the word Kandimalal is based on "*karnti*", the Jaru word for "yam" (*ibid*). Since these stories were not collected until after scientists investigated the structure, we cannot know with certainty the age of the stories.

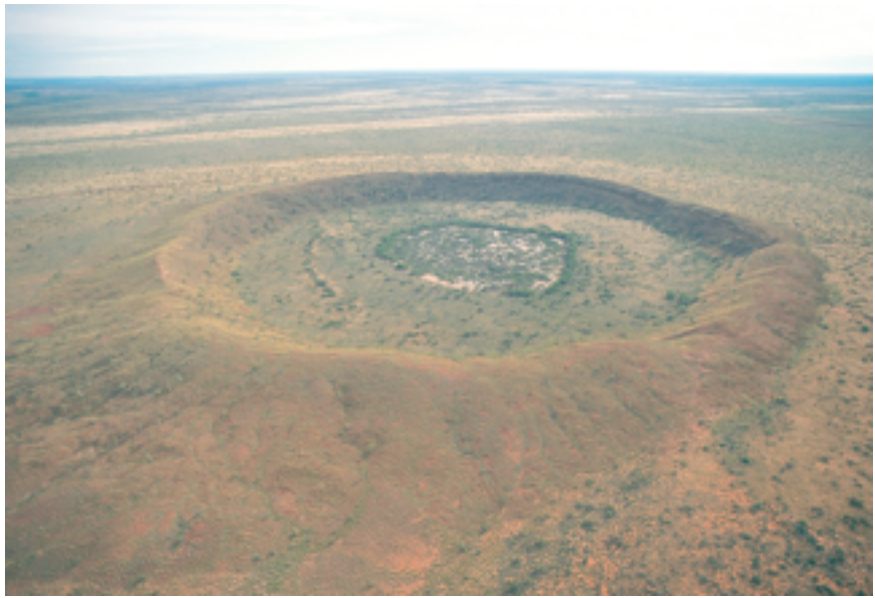

*Figure 6. An aerial view of Wolfe Creek crater. Photo by J. Goldsmith.*

Other elders recount stories that explain the structure's origin as an impact crater. Jaru elder Jack Jugarie (1927-1999) told a story of the crater's origin to the co-author (Goldsmith) in 1999, just prior to his passing. Jugarie said the story came from his grandfather's grandfather, suggesting that the account originated before the 1947 "discovery" of the crater. Jugarie clearly indicated that the story was "*according to the old people, early days mob, wild people that haven't seen gardia (white) people*". He explains that one night, the moon and the evening star passed very close to each other. The evening star became very hot and fell to the earth, causing a brilliant, deafening explosion. This greatly frightened the Jaru and it was a long time before they ventured near the site, only to discover it was the spot where the evening star had fallen (Goldsmith, 2000). Jugarie explains further:

> *"A star bin fall down. It was a small star, not so big. It fell straight down and hit the ground. It fell straight down and made that hole round, a very deep hole. The earth shook when that star fell down"* (Reeves-Sanday, 2007: 26).

Jugarie cites a Jaru word for a large meteor that caused the earth to shake - "*coolungmurru*". He said that after seeing a large meteor, the people would wait for the sound and feel the Earth shaking. Sonic effects caused by large exploding meteors (bolides) are well known to scientists, but are an uncommon phenomenon.

Jaru elder Stan Brumby (1933-2012; Figure 7) tells stories about Wolfe Creek Crater, particularly through his Aboriginal art. He features in a video exhibit at the Cosmology Gallery, Gravity Discovery Centre in Perth (Goldsmith, 2011) in which scientific and Aboriginal perspectives of Wolfe Creek Crater are shared. The video contains an interview with Brumby, recorded by Goldsmith while conducting ethnographic fieldwork:

> *"I sing him, that star, language, singing stick, I can sing 'im now. That's 'im song, Warda, Big star, bin fall down, from top from sky, Warda wandinga morunga."*

Aboriginal oral accounts relating to Wolfe Creek Crater are also closely related to contemporary Aboriginal artworks that feature the crater. Aboriginal artists who have represented Wolfe Creek crater are mainly based in Halls Creek, or at Billuluna (the nearest community to the crater). There are four main themes represented in these artworks. Firstly, the story of the star that fell to earth, forming the crater is shown in several of the paintings. Secondly, there are paintings that represent the crater itself, and essentially no other features, or the crater, the landforms and bush food in its general vicinity. Thirdly, some paintings show the story of the two rainbow snakes and the crater (as referred to in the National Park signage). One painting (Figure 8 – top left) shows a representation of a rainbow snake, coiled in the centre of the crater. Fourthly, some paintings show the belief of an underground tunnel, from the centre of the crater, which emerges at Sturt Creek, (at a place called Red Rock).

Jaru Elder Speiler Sturt (b. 1935) from Billiluna explains the cosmic origins of Wolfe Creek crater through story and illustration (Reeves-Sanday, 2007: 15; Figure 9):

> *"That star is a Rainbow Serpent. This is the Aboriginal Way. We call that snake Warnayarra. That snake travels like stars travel in the sky. It came down at Kandimalal. I been there, I still look for that crater. I gottem Ngurriny – that one,*

*Walmajarri/Jaru wild man."*

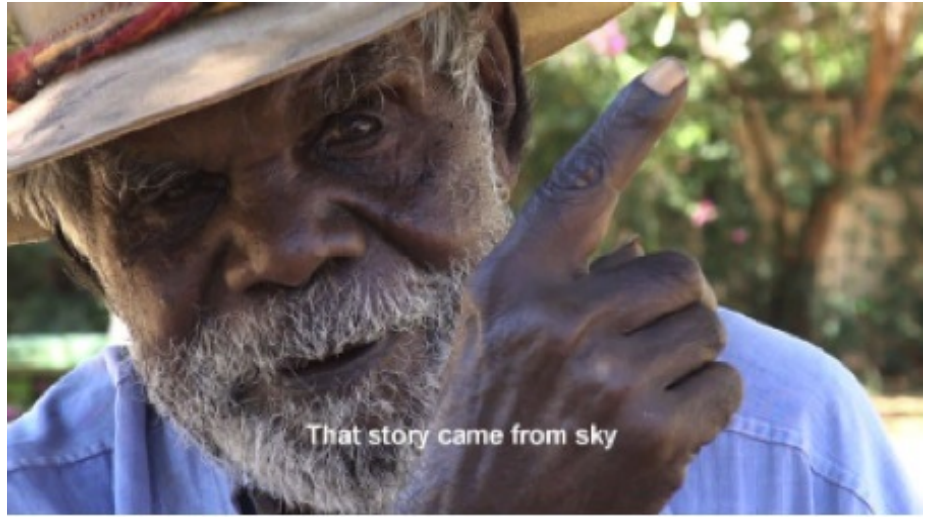

*Figure 7. Jaru elder Stan Brumby telling the story of Kandimalal. Photo by J. Goldsmith.*

Many recorded stories from Kandimalal describe the structure originating from a cosmic impact (see Reeves-Sanday, 2007). One such account presented here is said to originate prior to the 1947 Western discovery of the crater, however in general, it is difficult to determine whether this or similar accounts have been derived (or influenced by) the scientific explanations of the crater. In 2002, Walmajarri artist Jack Lannigan (b. 1924), a Jaru speaker, cited the story about the structure being formed from a giant serpent emerging from the ground (Reeves-Sanday, 2007: 97). When Reeves-Sanday asked him if the snake that formed the structure came from the sky, he replied "*nah*", the star-story was "*white-man's story*" (*ibid*: 99). The influence of the Western scientific explanation of the crater on local Aboriginal stories is evident. When artists developed paintings of the crater, they were encouraged to include the "star story" and were given directional advice about the theme.

In 1953, the anthropologist Norman Tindale (1953: 907-910) interviewed three Jaru men regarding stories related to the crater. According to the men, it was known as Kandimalal but they did not have any stories about it. From this, Tindale concluded that Kandimalal had no "special significance" to the Aboriginal people he interviewed. It is therefore unclear if the impact story predated the scientific rediscovery of the crater or if interest in the crater by scientists influenced the oral traditions. It seems that Aboriginal artists have indeed incorporated the impact story into their oral and artistic traditions. According to Reeves-Sanday (2007), the inclusion of the scientific story strengthened the power of the painting, revealing a willingness to embrace and incorporate new knowledge into pre-existing traditions.

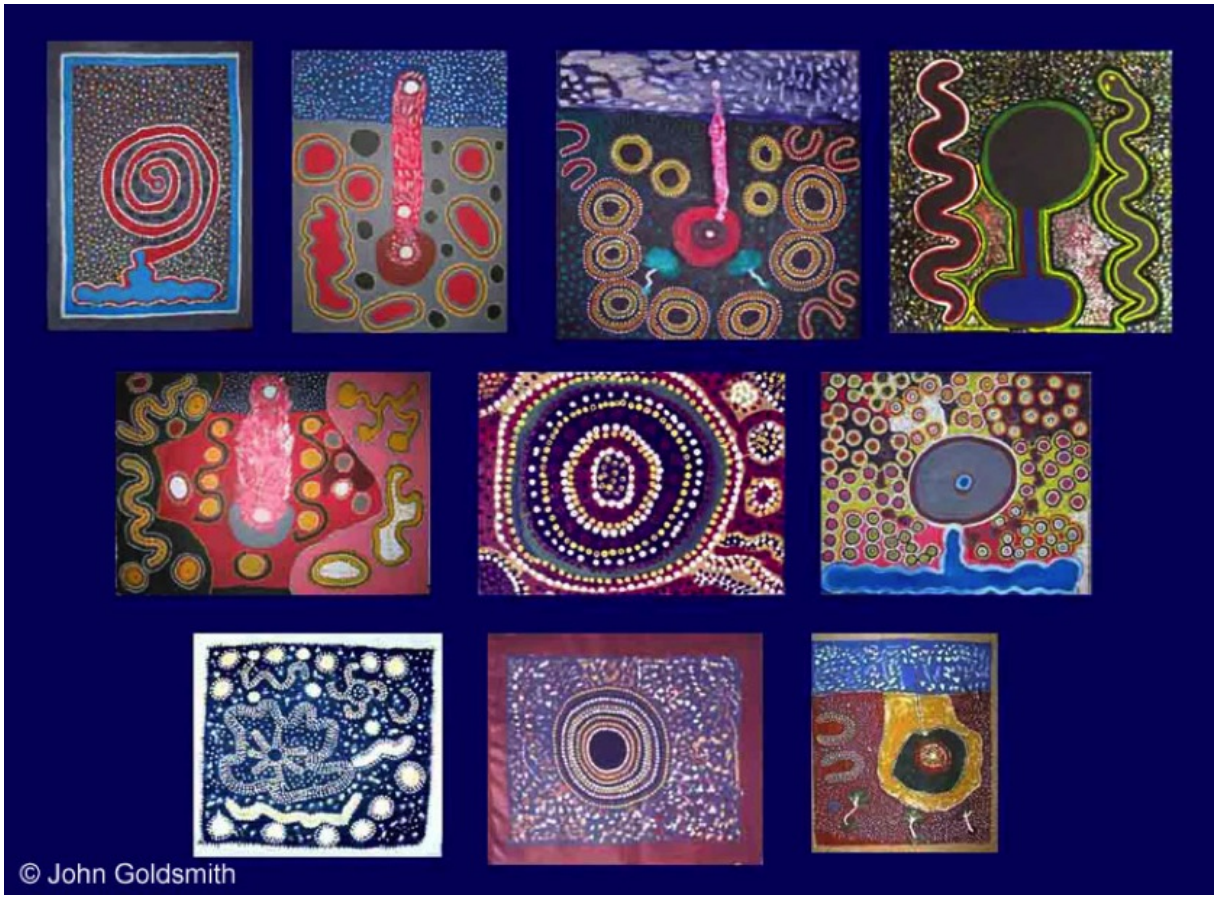

*Figure 8. Examples of Aboriginal artworks featuring Wolfe Creek Crater by Stan Brumby, Barbara Sturt, and Frank Clancy. Photo by J. Goldsmith. Reproduced by permission, Yarliyil Art Centre.*

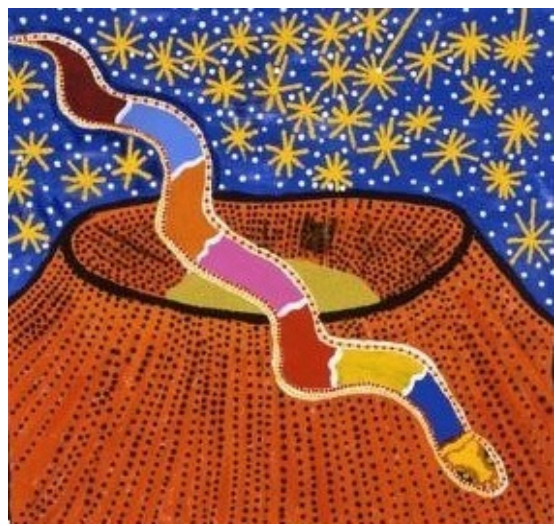

*Figure 9. Kandimalal (painting), by Speiler Sturt, 2003. Image used with permission (Reeves-Sanday 2007: Plate 16).*

3.      **Holocene Impact Craters with No Associated Oral Traditions**

3.1     **Boxhole Crater**

Boxhole crater, located ~170 km northeast of Alice Springs, is a simple impact structure approximately ~170 m in diameter. It was discovered by Joe Webb, a shearer at Boxhole Station. The age of the structure is uncertain, but estimates by Kohman and Goel (1963) using $^{14}$C exposure give 5400±1500 years while Shoemaker et al. (2005) and Haines (2005: 484-485) provide an estimate of ~30,000 years using $^{10}$Be/$^{26}$Al exposure.

Given the crater's geologically young age, it is possible Aboriginal people witnessed this event. This was apparent to Cecil T. Madigan (1889-1947), the first scientist to survey the crater on 20 June 1937:

> *"It seems hard to believe that the crater was formed before the country was occupied by the aborigines, yet such an explosion would surely have been noticed and recorded by them, whose lives are so uneventful and who are yet so observant of natural phenomena. It must have been audible, or visible at night, from a hundred miles away,"* (Madigan, 1937: 190).

He was surprised that local Aboriginal people seemed to have "*no legends connected with the crater*" and took "*no particular interest in it*", although he noted that Aboriginal people had dug a shallow soakage in the center of the crater where water collected. Webb and his nephews, who were in "*sympathetic contact with the Aborigines and masters of their language*", confirmed this.

A fragment of the Boxhole impactor was found in the Harts Range, 60 km south of Boxhole crater (de Laeter, 1973). de Laeter concluded that it was a transported fragment of the Boxhole impactor, most likely moved by Aboriginal people. Nothing more about the Aboriginal significance of this meteorite or the Boxhole crater is reported in the literature.

3.2     **Dalgaranga Crater**

Dalgaranga crater is one of the smallest impact structures in the world, one of the first to be identified in Australia, and the only terrestrial impact associated with a mesosiderite projectile (Nininger and Huss, 1960). The simple structure is located 100 km northeast of Yalgoo, Western Australia and has a diameter of 24 meters (Bevan, 1996). Although the age of the structure is not well known, its well-preserved nature suggests it might be as young as 3,000 years (Shoemaker and Shoemaker, 1988).

An Aboriginal stockman named Billy Seward discovered the crater in 1921 (Wellard, 1983: 95-97). Seward informed the station manager, G.E.P. Wellard, who returned to the site and found meteorite fragments in the area (*ibid*: 94-96). A study of the meteorites did not appear in the literature until 1938 (Simpson, 1938) and the crater was not surveyed until 1959 (Nininger and Huss, 1960). Details of the 1959 survey, which included Harvey Nininger, his wife, Addie, and amateur geologist Allen O. Kelly, were included in Kelly's unpublished memoirs (Kelly, 1961). Kelly identified flint flakes around the crater, which he thought might have marked places where meteorite fragments had been recovered by Aboriginal people as "*an exchange gift for the fiery god that came out of the sky*" (Kelly, 1961: 153-154). He speculated that the impact would have occurred within the last few hundred years and thought there was "little doubt" that Aboriginal people witnessed the impact. However, no Aboriginal stories related to the structure were identified in the literature and the crater's age has been estimated to have formed anywhere between 3,000 and 270,000 years ago (Shoemaker et al., 2005: 542).

A meteoritic slug found in Murchison Downs in 1925, some 200 km away, was identified as a fragment of the Dalgaranga impact (Bevan and Griffin, 1994). Like the Boxhole fragment, Bevan and Griffin suggest that Aboriginal people may have transported the slug from Dalgaranga. For a full history of research at Dalgaranga, see Hamacher and O'Neill (2013).

### 3.3    Veevers Crater

Veevers crater is a simple impact structure in the Canning Basin of central Western Australia, measuring 70 m in diameter (Haines, 2005). The crater was identified by the Bureau of Mineral Resources and the Geological Survey of Western Australia in 1975 and was confirmed as an impact structure in 1985 (Shoemaker and Shoemaker, 1985). The age of the crater is not well known and estimates range from < 20,000 years using cosmogenic nuclide exposure dating (Glikson, 1996) to < 4000 years based on the excellent preservation of the rim (Shoemaker and Shoemaker, 1988). No Aboriginal stories associated with the structure are reported in the literature.

### 4    Concluding Remarks

We find oral traditions associated with Henbury, Liverpool, and Wolfe Creek – three of the six youngest and smallest impact craters in Australia. Since many of the larger craters are either buried or eroded to such a point that they are not easily distinguishable from the surrounding landscape, we are unsurprised that we are unable to find stories associated with them. The only large (D > 2 km) crater with an associated oral tradition is Gosse's Bluff. Aboriginal oral traditions regarding Boxhole, Dalgaranga, and Veevers are not reported in the literature. As discussed before, we suspect that oral traditions associated with these and other craters may exist, but simply have not been collected or published. This is the focus of future work.

Of craters associated with reported Aboriginal knowledge, Gosse's Bluff, Henbury, and Wolfe Creek have oral traditions that relate to a cosmic impact, although it is uncertain if these stories predate colonisation or scientific investigations of the craters. The only oral tradition related to Henbury that is described in the anthropological literature does not relate to an impact event. However, Aboriginal traditions of Henbury in historical records indicates a living memory of the impact event. Oral traditions related to Liverpool are not associated with a cosmic impact. Both Henbury and Wolfe Creek crater are associated with both impact and non-impact oral traditions, although the influence of Western science on these traditions is evident, and confirmed by some Aboriginal elders for Wolfe Creek.

Aboriginal people seek explanations regarding the origin of natural features, including impact craters. These explanations, which are informed and influenced by new experiences and new information, are encoded in oral traditions that are passed through successive generations. This new information may be influenced by Western science. One Aboriginal informant claimed that his star story regarding Wolfe Creek Crater originated before contact with white people, whereas some informants stated that views from Western science were incorporated into their traditions. We do not know if this has occurred with respect to Gosse's Bluff or Henbury craters, but we do know that Carolyn Shoemaker discussed the origin of Liverpool crater with Aboriginal custodians. It would be interesting to know if contemporary Aboriginal traditions now incorporate the information provided by Shoemaker.

The Aboriginal accounts of the Henbury craters suggest that the event has survived in living memory after more than 4,500 years. However, we are unable to definitively demonstrate that Aboriginal traditions of meteorite craters existed prior to colonisation or scientific investigation that describe them as having an impact origin. Interest in the Henbury craters in the local region existed for decades before it was confirmed as an impact crater and it is unknown if and how this interest influenced Aboriginal traditions of the site. Similarly, scientific interest in Gosses Bluff may have influenced local Aboriginal traditions, but we cannot provide evidence of this in the literature. Instead, we must rely on the Aboriginal custodians of the stories themselves or on historical accounts, such as that of Henbury. It must be emphasised that contemporary traditions are just as important as "pre-colonisation" traditions, whether they incorporate Western knowledge or not. Contemporary Indigenous traditions of impact craters have relevance to the identity and spirituality of modern Aboriginal people.

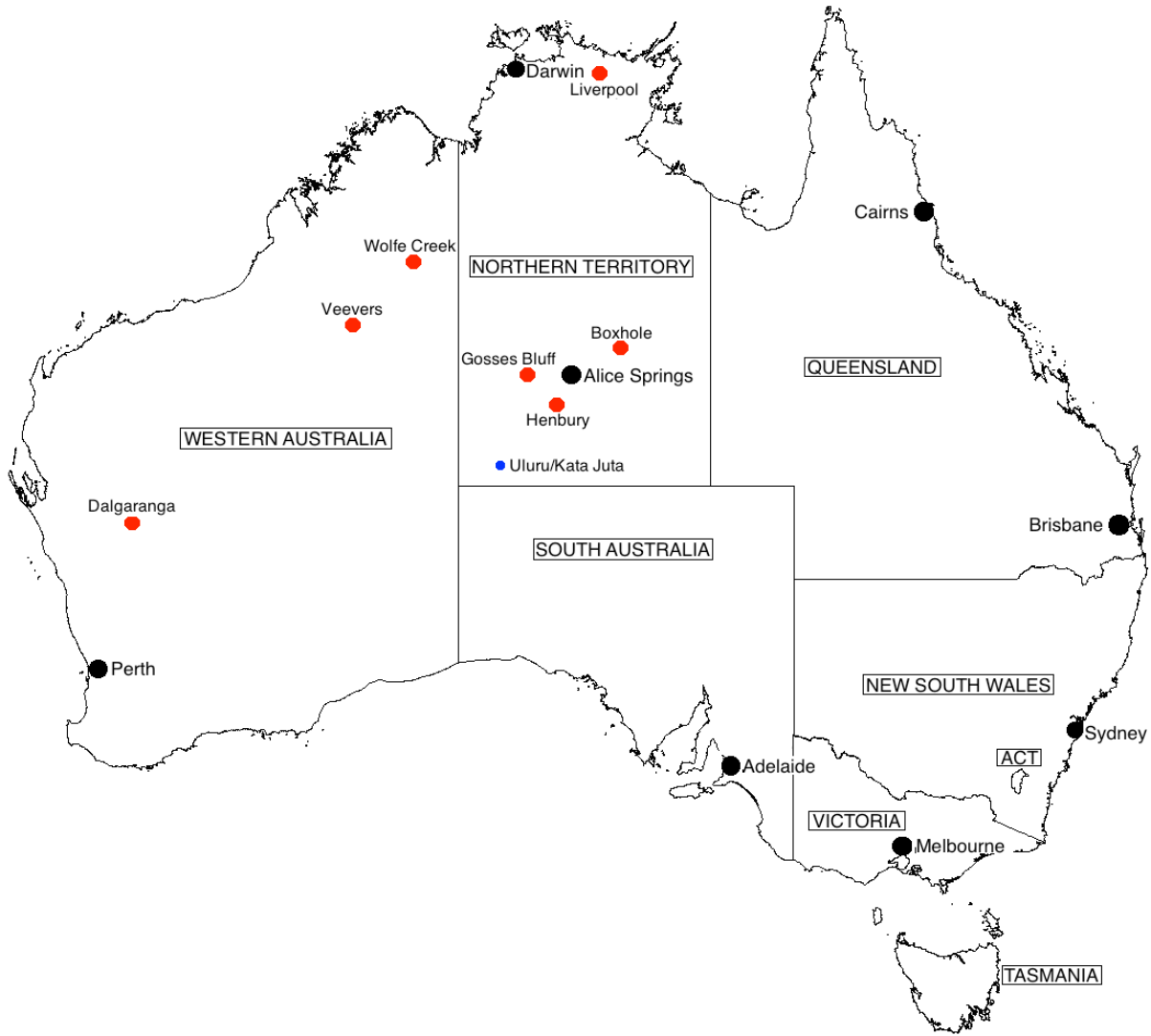

*Figure 10: A map of places described in this paper. Craters described in the text are indicated in red.*

Additional ethnographic and archaeological research at these sites is required and is the subject of future work. The co-author (Goldsmith) has conducted ethnographic fieldwork at Wolfe Creek crater, which is the topic of his doctoral thesis and will be published in the future.

Aboriginal knowledge regarding impact craters continues to be developed though art and storytelling, especially in regards to Wolfe Creek crater. Art and story are important ways for contemporary Aboriginal people to share and preserve this knowledge for future generations and we hope that future work will expand on Aboriginal oral traditions of impact sites for the Aboriginal and scientific communities.

**Acknowledgements**

We wish to acknowledge the traditional custodians of the places discussed in this paper and pay our respects to Mavis Malbunka, Speiler Sturt, Stan Brumby, Johnny Maurirundjul, Jimmy Njimimjuna, Barbara Sturt, Jack Jugarie, Jack Lannigan, and Frank Clancy. We also wish to thank Jeanette Swan, Yarilyil Art Centre (Halls Creek), Craig O'Neill, Peggy Reeves-Sanday, Don McColl, Svend Buhl, Jason Heffernan, Jason Wright, Steve Hutcheon, and David Nash for their advice and assistance. Some of the data and photos were collected during the Meteoritical Society 2012 post-conference fieldtrip to the Central Desert. This work also made use of the NASA Astrophysics Data System, JSTOR, and Trove databases.

**Notes**

1. An interesting note relevant to the Henbury research is taken from Spencer and Gillen (1899: 549), before the craters were known to Westerners. In some Central Desert groups, the Aboriginal people believed in a form of evil magic called Arungquiltha. This magic sometimes took the form of a meteor streaking across the sky "like a ball of fire". In the Central Desert, mushrooms were believed to come from falling stars and contained Arungquiltha. They were considered taboo and their consumption was forbidden (Spencer and Gillen, 1904: 627). According to Spencer and Gillen, a man "out west" was found dead and mutilated. The suspected perpetrators were certain men living at Henbury on the Finke River, who were are accused of projecting the Arungquiltha. The Henbury craters are just 7 km south of the Finke River (Spencer and Gillen, 1899: 549).

2. Minor elements of this story are deliberately excluded since they are considered secret and are not pertinent to the paper. Although Mountford collected this story from a male informant, one of the authors (Hamacher) was asked to leave out certain elements of the story, as they are considered sensitive.

**About the Authors**

Duane Hamacher is a Lecturer in the Nura Gili Indigenous Centre at the University of New South Wales in Sydney, Australia. He publishes extensively on cultural and historical astronomy and meteoritics, with a focus on Indigenous Australia. He earned BSc and MSc degrees in Astrophysics and completed a PhD with a thesis entitled "*On the astronomical knowledge and traditions of Aboriginal Australians*" at Macquarie University in Sydney. He is a consultant curator and astronomy educator at Sydney Observatory and the founder and Chair of the Australian Society for Indigenous Astronomy.

John Goldsmith is a researcher and astrophotographer in Perth, Western Australia. He earned BSc and MSc degrees in Environmental Science and completed a PhD with a thesis entitled "*Cosmos, Culture, and Landscape: Documenting, Learning and Sharing Aboriginal Astronomical Knowledge in Contemporary Society*" at Curtin University in Perth. He chaired and coordinated the development of the Cosmology Gallery, Gravity Discovery Centre (Gingin WA), is the curator of the WA Astrofest Astrophotography exhibition (since 2009), and is the creative developer behind the "Celestial Visions" exhibition. He is a member of a global astrophotography network, "The World At Night", and in 2011 was the inaugural tertiary recipient of the 2011 de Laeter Science Engagement Scholarship, Gravity Discovery Centre Foundation.